# Modeling the triple-GEM detector response to background particles for the CMS Experiment

**M. Abbas,[n] M. Abbrescia,[t] H. Abdalla,[h,j] A. Abdelalim,[h,k] S. AbuZeid,[h,i] A. Agapitos,[d] A. Ahmad,[af] A. Ahmed,[q] W. Ahmed,[af] C. Aimè,[y] C. Aruta,[t] I. Asghar,[af] P. Aspell,[ak] C. Avila,[f] I. Azhgirey,[ag] J. Babbar,[p] Y. Ban,[d] R. Band,[am] S. Bansal,[p] L. Benussi,[v] V. Bhatnagar,[p] M. Bianco,[ak] S. Bianco,[v] K. Black,[ap] L. Borgonovi,[u] O. Bouhali,[al] A. Braghieri,[y] S. Braibant,[u] S. Butalla,[aq] S. Calzaferri,[y] M. Caponero,[v] F. Cassese,[x] A. Castaneda,[ae,1] N. Cavallo,[x] S. S. Chauhan,[p,2] A. Colaleo,[t] A. Conde Garcia,[ak] M. Dalchenko,[al] A. De Iorio,[x] G. De Lentdecker,[a] D. Dell Olio,[t] G. De Robertis,[t] W. Dharmaratna,[aj] S. Dildick, [al,3] B. Dorney,[a] R. Erbacher,[am] F. Fabozzi,[x] F. Fallavollita,[ak] A. Ferraro,[y] D. Fiorina,[y] E. Fontanesi,[u] M. Franco,[t] C. Galloni,[ap] P. Giacomelli,[u] S. Gigli,[y] J. Gilmore,[al] M. Gola,[q] M. Gruchala,[ak] A. Gutierrez,[an] R. Hadjiiska,[c] T. Hakkarainen,[l] J. Hauser,[ao] K. Hoepfner,[m] M. Hohlmann,[aq] H. Hoorani,[af] T. Huang,[al] P. Iaydjiev,[c] A. Irshad,[a] A. Iorio,[x] F. Ivone,[m] J. Jaramillo,[g] V. Jha,[s] A. Juodagalvis,[ad] E. Juska,[al] B. Kailasapathy,[ah,ai] T. Kamon,[al] Y. Kang,[ab] P. Karchin,[an] A. Kaur,[p] H. Kaur,[p] H. Keller,[m] H. Kim,[al] J. Kim,[aa] S. Kim,[ab] B. Ko,[ab] A. Kumar,[q] S. Kumar,[p] H. Kumawat,[s] N. Lacalamita,[t] J.S.H. Lee,[ab] A. Levin,[d] Q. Li,[d] F. Licciulli,[t] L. Lista,[x] K. Liyanage,[aj] F. Loddo,[t] M. Luhach,[p] M. Maggi,[t] Y. Maghrbi[ac] N. Majumdar,[r] K. Malagalage,[ah] S. Malhotra,[al] S. Mallows,[n] S. Martiradonna,[t] N. Mccoll,[ao] C. McLean,[am] J. Merlin,[t] M. Misheva,[c] D. Mishra,[s] G. Mocellin,[m] L. Moureaux,[a] A. Muhammad,[af] S. Muhammad,[af] S. Mukhopadhyay,[r] M. Naimuddin,[q] P. Netrakanti,[s] S. Nuzzo,[t] R. Oliveira,[ak] L. Pant,[s] P. Paolucci,[x] I.C. Park,[ab] L. Passamonti,[v] G. Passeggio,[x] A. Peck,[ao] A. Pellecchia,[t] N. Perera,[aj] L. Petre,[a] H. Petrow,[l] D. Piccolo,[v] D. Pierluigi,[v] G. Raffone,[v] M. Rahmani,[aq] F. Ramirez,[g] A. Ranieri,[t] G. Rashevski,[c] B. Regnery,[am] M. Ressegotti, [y,4] C. Riccardi,[y] M. Rodozov,[c] E. Romano,[y] C. Roskas,[b] B. Rossi,[x] P. Rout,[r] D. Roy,[aq] J. D. Ruiz,[g] A. Russo,[v] A. Safonov,[al] A. K. Sahota,[p] D. Saltzberg,[ao] G. Saviano,[v] A. Shah,[q] A. Sharma,[ak] R. Sharma,[q] T. Sheokand,[p] M. Shopova,[c] F. Simone,[t] J. Singh,[p] U. Sonnadara,[ah] E. Starling,[a] B. Stone,[ao] J. Sturdy,[an] G. Sultanov,[c] Z. Szillasi,[o] D. Teague,[ap] D. Teyssier,[o] T. Tuuva,[l] M. Tytgat,[b] I. Vai[w] N. Vanegas,[g] R. Venditti,[t] P. Verwilligen,[t] W. Vetens,[ap] A. K. Virdi,[p] P. Vitulo,[y] A. Wajid,[af] D. Wang,[d] K. Wang,[d] I. J. Watson,[ab] N. Wickramage,[aj] W. Jang,[ab] D.D.C. Wickramarathna,[ah] S. Yang,[ab] Y. Yang,[a] U. Yang,[aa] J. Yongho,[z] I. Yoon,[aa] Z. You,[e] I. Yu[z] and S. Zaleski[m] on behalf of the CMS Collaboration**

[a] *Université Libre de Bruxelles, Bruxelles, Belgium*

[b] *Ghent University, Ghent, Belgium*

[c] *Institute for Nuclear Research and Nuclear Energy, Bulgarian Academy of Sciences, Sofia, Bulgaria*

[1] Corresponding authors.

[2] Corresponding authors.

[3] Now at Rice University, Houston, Texas, USA

[4] Now at INFN Sezione di Genova, Genova, Italy

[d] *Peking University, Beijing, China*
[e] *Sun Yat-Sen University, Guangzhou, China*
[f] *University de Los Andes, Bogota, Colombia*
[g] *Universidad de Antioquia, Medellin, Colombia*
[h] *Academy of Scientific Research and Technology - ENHEP, Cairo, Egypt*
[i] *Ain Shams University, Cairo, Egypt*
[j] *Cairo University, Cairo, Egypt*
[k] *Helwan University, also at Zewail City of Science and Technology, Cairo, Egypt*
[l] *Lappeenranta University of Technology, Lappeenranta, Finland*
[m] *RWTH Aachen University, III. Physikalisches Institut A, Aachen, Germany*
[n] *Karlsruhe Institute of Technology, Karlsruhe, Germany*
[o] *Institute for Nuclear Research ATOMKI, Debrecen, Hungary*
[p] *Panjab University, Chandigarh, India*
[q] *Delhi University, Delhi, India*
[r] *Saha Institute of Nuclear Physics, Kolkata, India*
[s] *Bhabha Atomic Research Centre, Mumbai, India*
[t] *Politecnico di Bari, Università di Bari and INFN Sezione di Bari, Bari, Italy*
[u] *Università di Bologna and INFN Sezione di Bologna, Bologna, Italy*
[v] *Laboratori Nazionali di Frascati INFN, Frascati, Italy*
[x] *Università di Napoli and INFN Sezione di Napoli, Napoli, Italy*
[y] *Università di Pavia and INFN Sezione di Pavia, Pavia, Italy*
[w] *Università di Bergamo and INFN Sezione di Pavia, Pavia, Italy*
[z] *Korea University, Seoul, Korea*
[aa] *Seoul National University, Seoul, Korea*
[ab] *University of Seoul, Seoul, Korea*
[ac] *College of Engineering and Technology, American University of the Middle East, Dasman, Kuwait*
[ad] *Vilnius University, Vilnius, Lithuania*
[ae] *Universidad de Sonora, Hermosillo, Mexico*
[af] *National Center for Physics, Islamabad, Pakistan*
[ag] *Institute for High Energy Physics of NRC Kurchatov Institute, Protvino, Russia*
[ah] *University of Colombo, Colombo, Sri Lanka*
[ai] *Trincomalee Campus, Eastern University, Sri Lanka, Nilaveli, Sri Lanka*
[aj] *University of Ruhuna, Matara, Sri Lanka*
[ak] *CERN, Geneva, Switzerland*
[al] *Texas A&M University, College Station, USA*
[am] *University of California, Davis, Davis, USA*
[an] *Wayne State University, Detroit, USA*
[ao] *University of California, Los Angeles, USA*
[ap] *University of Wisconsin, Madison, USA*
[aq] *Florida Institute of Technology, Melbourne, USA*

*E-mail:* schauhan@cern.ch, castaned@cern.ch

Abstract: An estimate of environmental background hit rate on triple-GEM chambers is performed using Monte Carlo (MC) simulation and compared to data taken by test chambers installed in the CMS experiment (GE1/1) during Run-2 at the Large Hadron Collider (LHC). The hit rate is measured using data collected with proton-proton collisions at 13 TeV and a luminosity of $1.5\times10^{34}$ cm$^{-2}$ s$^{-1}$. The simulation framework uses a combination of the FLUKA and Geant4 packages to obtain the hit rate. FLUKA provides the radiation environment around the GE1/1 chambers, which is comprised of the particle flux with momentum direction and energy spectra ranging from $10^{-11}$ to $10^{4}$ MeV for neutrons, $10^{-3}$ to $10^{4}$ MeV for $\gamma$'s, $10^{-2}$ to $10^{4}$ MeV for $e^{\pm}$, and $10^{-1}$ to $10^{4}$ MeV for charged hadrons. Geant4 provides an estimate of detector response (sensitivity) based on an accurate description of detector geometry, material composition and interaction of particles with the various detector layers. The MC simulated hit rate is estimated as a function of the perpendicular distance from the beam line and agrees with data within the assigned uncertainties of 10-14.5 %. This simulation framework can be used to obtain a reliable estimate of background rates expected at the High Luminosity LHC.

## Contents

## 1 Introduction

Modern particle physics collider experiments use collimated high energy particle beams to either collide them or dump on a fixed target. Due to the high collision rate and the interaction of beam particles with matter a hostile radiation environment is created. This radiation field is composed mainly of low energy neutrons, photons ($\gamma$), electrons/positrons ($e^{\pm}$) and charged hadrons namely kaon ($K^{\pm}$), pion ($\pi^{\pm}$), proton ($p^{+}$) [1]. Particles in the radiation field are commonly known as background particles and they have an energy typically below 1 GeV. Due to the large interaction cross-section those background processes can cause damage to detector elements and front-end electronics [2], additionally they can induce spurious signals that degrade detector performance. A strategy for radiation protection is crucial during the design and upgrade of LHC experimental facilities [3]. This strategy uses estimates obtained with Monte Carlo (MC) simulation and experimental measurements. Dedicated data taking campaigns are used to collect background data to understand detector behavior with respect to variations in experimental parameters such as luminosity and detector coordinates (i.e. R, z, $\phi$). Additionally new detector technologies are studied using high radiation doses in dedicated facilities, for instance the CERN High Energy Accelerator Mixed Field (CHARM) [4] and the Gamma Irradiation Facility (GIF++) [5] at CERN. A set of detectors based on triple-GEM technology [6] is being installed in the Compact Muon Solenoid (CMS) detector for Run-3 and Phase-2 (High Luminosity LHC) muon upgrade program [7] and similar technology has been adopted for the upgrade of the ALICE TPC [8]. Simulation plays a key role in understanding the impact of radiation on detector performance, predict behavior with an increased radiation dose and prepare strategies for mitigation. To study the radiation environment and its impact on detector performance dedicated simulation tools are used, those involve a description of complex physics processes and particle decay chains. The present study extends our previous work [7, 9] by discussing in depth the simulation strategy and methodology, improvement in the detector description and the estimation of detector response for a single triple-GEM detector as a function of incident particle's kinematical properties (energy, momentum direction). This work could also be useful for other muon systems simulation studies to estimate background response [10, 11].

The present study aims to serve as a reference for detector response to background particles, including a detailed description of the simulation strategy and methodology, estimation of detector response for a single triple-GEM detector as a function of incident particle's kinematical properties (energy, momentum direction). Variations to detector configuration are considered to show the validity of the estimations and robustness of the simulation model. Furthermore, the simulation model is compared with experimental data collected during 2018 by the GE1/1 slice test exercise at the CMS detector. Having a reliable modeling of the detector response will serve for future studies which are relevant considering the preparation for the High Luminosity LHC [12] in which the collision rate will increase one order of magnitude and the different detector technologies will experience a radiation environment never seen before.

## 2 Response of a single triple-GEM chamber to environmental background particles

### 2.1 Single triple-GEM geometry

The detector response is modeled using a Geant4 [13] simulation framework with the geometry of a triple-GEM detector [7] and incident background particles with properties consistent with those generated by the proton-proton collisions at the LHC. The dimensions and material composition for a single triple-GEM detector in figure 1 and table 1 approximates the prototype described in the GE1/1 detector technical report [7]. A complete description of the CMS GE1/1 chambers with readout chips and cooling system are given in section 3.2. The coordinate system adapted by CMS is a right-handed cartesian coordinate system with the origin at the collision point, the $x$-axis pointing towards the center of the LHC ring, the y-axis pointing upwards and $z$-axis along the beam direction. The polar angel $\theta$ is measured from the positive $z$-axis to the $x$-$y$ plane and the azimuthal angle $\phi$ is measured from the positive $x$-axis in the $x$-$y$ plane.

### 2.2 Simulation

The physics processes and decay chains are modeled using Geant4 version 10.6 and the recommended physics list for standard HEP processes (FTFP_BERT_HP). It includes all standard electromagnetic processes, Bertini-style cascade for hadrons (< 5 GeV), FTF (Fritiof) model for high energies (>4 GeV). It also includes a high precision neutron model used for neutrons below 20 MeV [14].

The simulation setup consist of a source plane, same size as drift board of GEM detector, for primary particle generation at a distance of 3 mm from the surface of the detector on both side on the $z$-axis. Such close placement of the source on the triple-GEM detector captures all possible angles of primary particles hitting the surface as well as all possible length of primary inside the sensitive volume of detector. The response of the detector for primary incident particle is measured separately for each source plane on either side individually and then the average is estimated as final response of the detector. Although the simulation considers the interaction of particles and creation of secondaries through the different layers in the triple-GEM, the detector response is extracted from the first two gas gaps namely drift and transfer-1 gap.

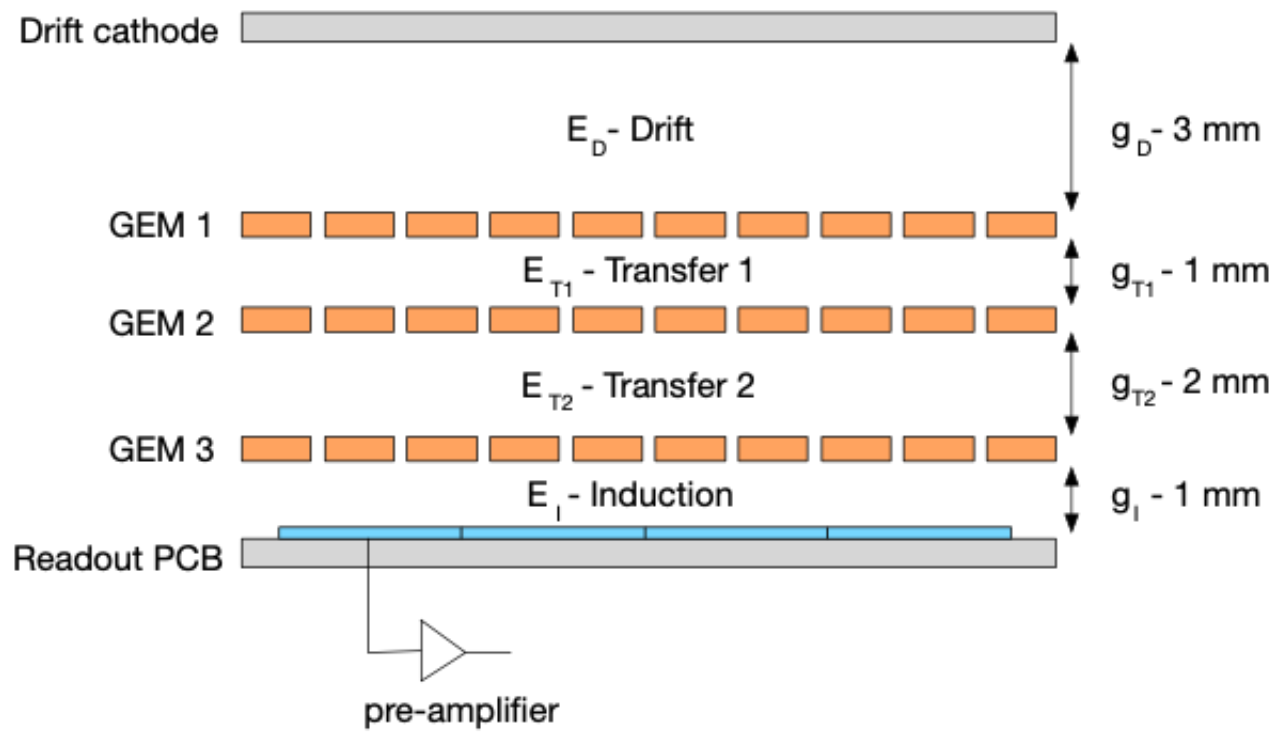

**Figure 1**. Representation of a transversal view of a triple-GEM detector.

**Table 1**. Material and dimensions of the different layers in a single triple-GEM detector as per CMS coordinate system.

| Layer | z-Dimensions | Material |
|---|---|---|
| Drift Board | 35 $\mu$m/3.2 mm/35 $\mu$m | Copper/FR4/Copper |
| Drift Gap | 3 mm | $Ar/CO_2$ |
| GEM1 | 5 $\mu$m/50 $\mu$m/5 $\mu$m | Copper/Kapton/Copper |
| Transfer 1 Gap | 1 mm | $Ar/CO_2$ |
| GEM2 | 5 $\mu$m/50 $\mu$m/5 $\mu$m | Copper/Kapton/Copper |
| Transfer 2 Gap | 2 mm | $Ar/CO_2$ |
| GEM3 | 5 $\mu$m/50 $\mu$m/5 $\mu$m | Copper/Kapton/Copper |
| Induction Gap | 1 mm | $Ar/CO_2$ |
| Readout Board | 35 $\mu$m/3.2 mm/35 $\mu$m | Copper/FR4/Copper |

This signal induction is due to the production of a charged particle due to conversion of neutral background particles (neutrons or $\gamma$'s) or direct interaction from electrons and charged hadron coming from outside this region. The rest of the processes in the signal evolution, electron drift, multiplication, charge transfer and electronic response are not covered in this study as those are usually considered for dedicated studies on optimization of signal detection [15], in which other simulation packages are used.

Detector response is evaluated using the *sensitivity* variable. *Sensitivity* is defined as the probability for a charged particle to deposit energy in the sensitive volume (e.g., $Ar/CO_2$ gas), producing primary ionized electrons [7, 9]. Those electrons are multiplied, so the charge is large enough to be detected by a readout system with charge thresholds. Thus the charge threshold is related to the energy deposit required to separate noise from signal. The simulated energy deposited in the first two gaps of a triple-GEM detector is used to set the minimum charge deposition required to emulate the front-end electronics discriminator value used to suppress noise hits. The values are chosen to correspond to the minimum production of one electron in the drift gap and 24 electrons in the transfer-1 gap and are set to 28.1 eV and 674.4 eV per incident particle assuming a gain of approximately $\sim 10^4$. Later in section 3.2 the energy threshold is set to match the operational configuration of the GE1/1 chambers at CMS. It should be emphasized here that instead of counting secondary charge particles in the gaps we require energy deposits which is equivalent to requiring minimum one simulated hit in the gaps and is also used in ref. [9]. The counting of secondary charges is not used due to the fact that Geant4 only reports them if they pass the default production cut threshold of 0.7 mm for track length [16].

According to the type of particle, their kinematical properties, and the material composition in which the particle interacts with, the following processes are the dominant ones: for high energy neutrons (> 10 MeV) the dominant process is inelastic scattering, while for intermediate ($10^{-5}$ to 10 MeV) and low energy (< $10^{-5}$ MeV) are the elastic scattering and neutron capture respectively. For high energy $\gamma$'s (> 10 MeV) the pair production is dominant, while for intermediate ($10^{-1}$ - 10 MeV) and low values ($10^{-1}$ MeV) are the Compton scattering and photoelectric effect respectively. In case of high energy $e^{\pm}$ (> 1 MeV) Bremsstrahlung radiation has a larger contribution while for low energy (< 1 MeV) ionisation process is dominant. Neutral particles (neutrons and $\gamma$'s) need to undergo a conversion reaction before they produce a charged particle. The sensitivity estimation as a function of type of incident particle, kinetic energy and angle is presented as a 2D matrix in figures 2 left, figure 2 right for neutrons and $\gamma$'s. The one dimensional projection (sensitivity as a function of incident energy) for different particles and angle perpendicular to the detector surface is presented at figure 3 left, the sensitivity as a function of incident angle for two energy values (1 and 100 MeV) is presented in figure 3 right.

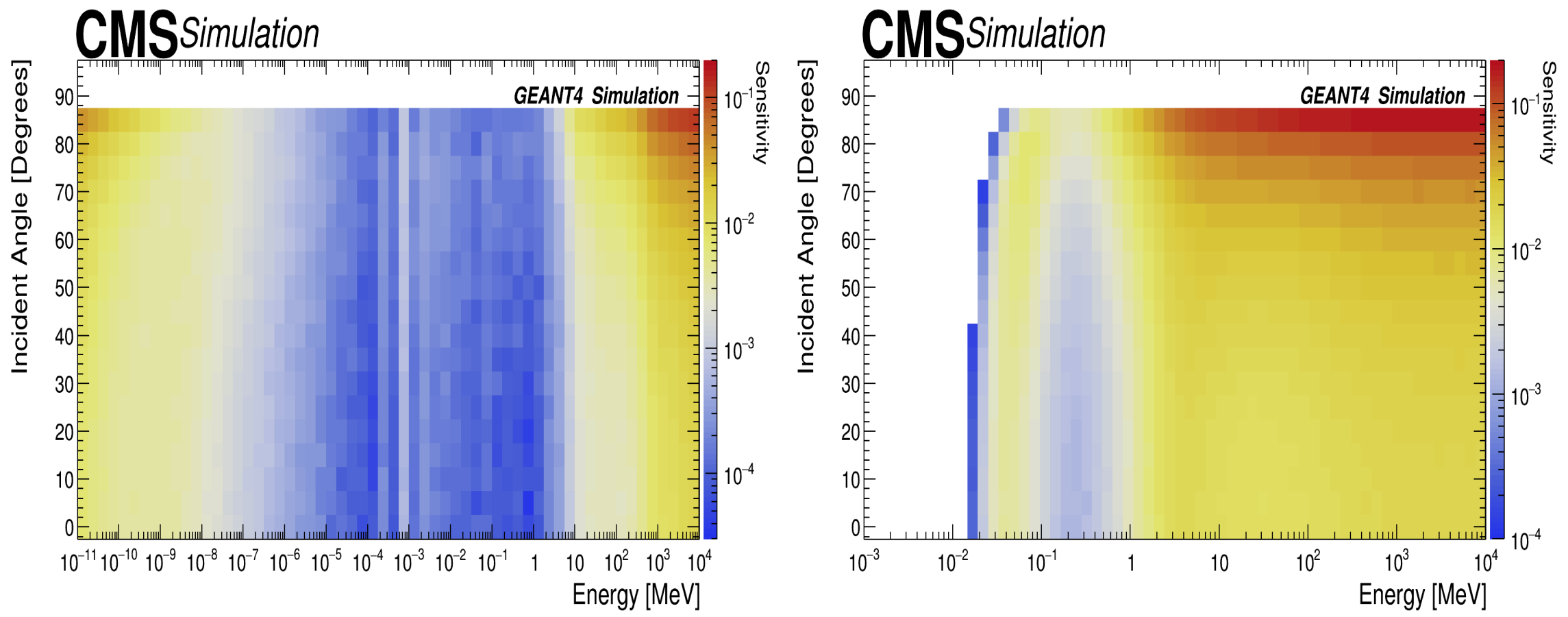

**Figure 2**. Sensitivity map for neutrons (left) and photon (right); in the x-axis is the kinetic energy, y-axis is the incident angle, the sensitivity value is presented in the color band.

Sensitivity has a strong dependence on the dynamics of incident particles, in particular this is true for the kinetic energy and incident angle. The probability of interaction is directly correlated to the detector width and the incidence angle: if the incidence angle increases the particle will traverse a slightly larger distance inside the detector, which means more material is available as target for the interaction. On the other hand if the kinetic energy increases, and depending on the nature of the incident particle, the probability to interact also increases according to the characteristic cross section for each type of particle. Results were obtained using large simulated samples $O(10^7)$ events, statistical uncertainty variations depends on the type of particle under study and the region on the sensitivity map, with the following ranges; 0-16% (neutrons), 0-2.1 % ($\gamma$'s) and 0-10.0 % ($e^{\pm}$).

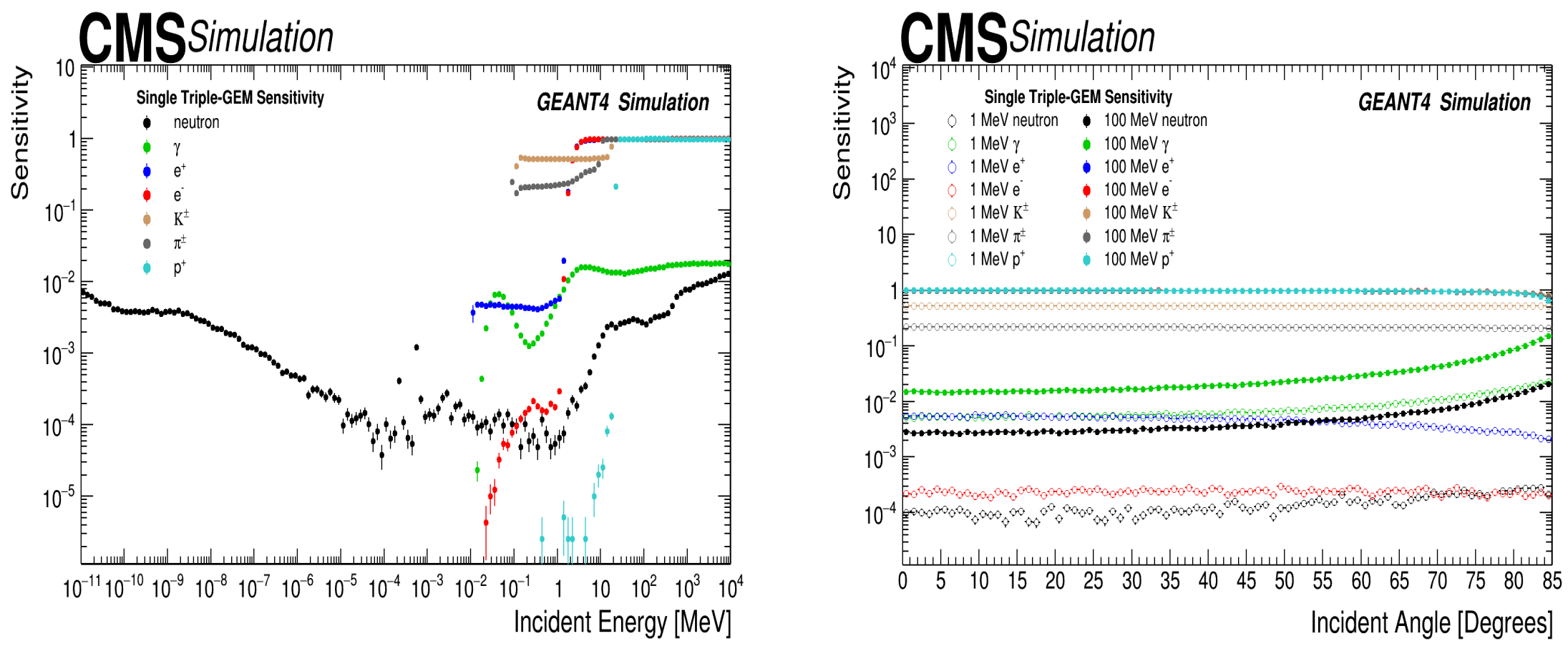

**Figure 3**. Sensitivity as a function of kinetic energy, for neutrons, photons and electrons, considering an incident angle of zero degrees with respect to normal to the detector surface (left). Sensitivity as a function of incident angle for different type of particles considering two energy values of 1 MeV and 100 MeV (right).

It is important to notice that the sensitivity results were obtained without constraints on particular radiation environment, the only information considered was the energy range for the different type of particles. This means that the results can be used in any experimental facilities in which similar triple-GEM detectors are installed [9]. The results presented in figure 3 left for single triple-GEM detector with basic geometry are qualitatively similar and quantitatively compatible with those presented in ref. [9]. The main difference is observed in neutron sensitivity at low energies which is largely attributed to the fact that Geant4 version in ref. [9] is not the same and threshold for drift and transfer-1 gap were not applied. It should also be noted here that a single triple-GEM detector response presented in figure 3 is generic and will be different quantitatively once CMS specific details and setup of the GEM detectors geometry are included. In the following sections the background modeling is adapted to the characteristic radiation environment of the CMS experiment and the specific detector geometry of the GE1/1 muon upgrade project.

## 3 Environmental background particles on the GE1/1 chambers at CMS

Five superchambers were installed (discussed in detail in section 3.2) during 2017-2018, as shown in figure 4, in an exercise known as "slice test", whose aim was to test electronic integration and experience in situ in the CMS cavern during LHC running conditions.

To recreate the experimental conditions and to build a simulation model, two components need to be considered, the first one is the simulation of the radiation environment. The radiation environment is estimated using a FLUKA [17] simulation with the experimental configuration of the CMS experiment at the LHC in which two proton beams collide with a center of mass energy of 13 TeV. The simulation models the particle interactions and transport through the different layers of the CMS detector. The second component is the detector response following a similar approach to the method described in section 2 but using the geometry of a superchamber. Both components are described in the following sections.

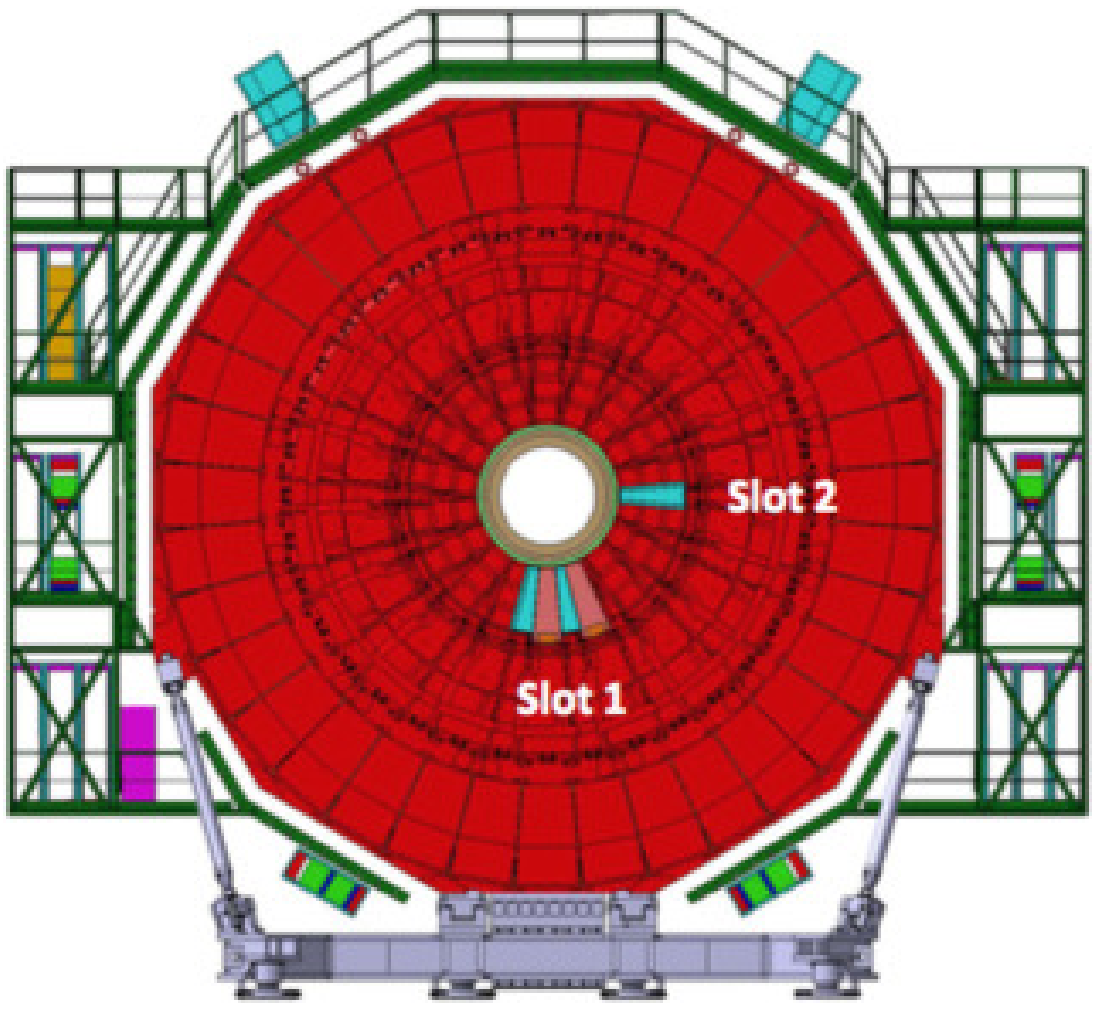

**Figure 4**. Schematic drawing of the negative muon endcap, showing the location of the five slice test superchambers.

### 3.1 Radiation Environment for GE1/1

The information for incoming particles reaching GE1/1 system is extracted from a FLUKA simulation with the parameters set assuming the CMS data taking conditions for Run-2 i.e. a centre of mass energy of $\sqrt{s} = 13$ TeV and a luminosity of $1.5 \times 10^{34}$ cm$^{-2}$ s$^{-1}$. A cross-section of the CMS detector geometry used in the FLUKA is shown in figure 5 top-left. One of the main interesting variables from the FLUKA simulation is the flux of particles, which represent the number of particles per unit area and per unit time, for the volume defined by the detector of interest; such a distribution is presented in figure 5 top-right. Other relevant distributions to characterize incoming particles are the energy spectra and incident angle which are variables, needed later for the Geant4 simulation, such variables are presented in figures 5 bottom-left and 5 bottom-right respectively.

According to FLUKA simulation [18] the major contribution in the region where the GE 1/1 detectors are located comes from low energy neutrons ($10^{-11}$ to $10^4$ MeV), $\gamma$'s ($10^{-3}$ to $10^4$ MeV), $e^{\pm}$ ($10^{-2}$ to $10^4$ MeV) and charged hadrons ($10^{-1}$ to $10^4$ MeV). This composition dominates the background mixed field for LHC experiments, other minor contributions and prompt particles are neglected as their production cross section is order of magnitude lower than those of background particles and therefore are not relevant for the purpose of this study. Prompt particles are usually studied using dedicated frameworks which are not suitable for the simulation of long-lived particles, such as neutrons, due to the short simulation time window.

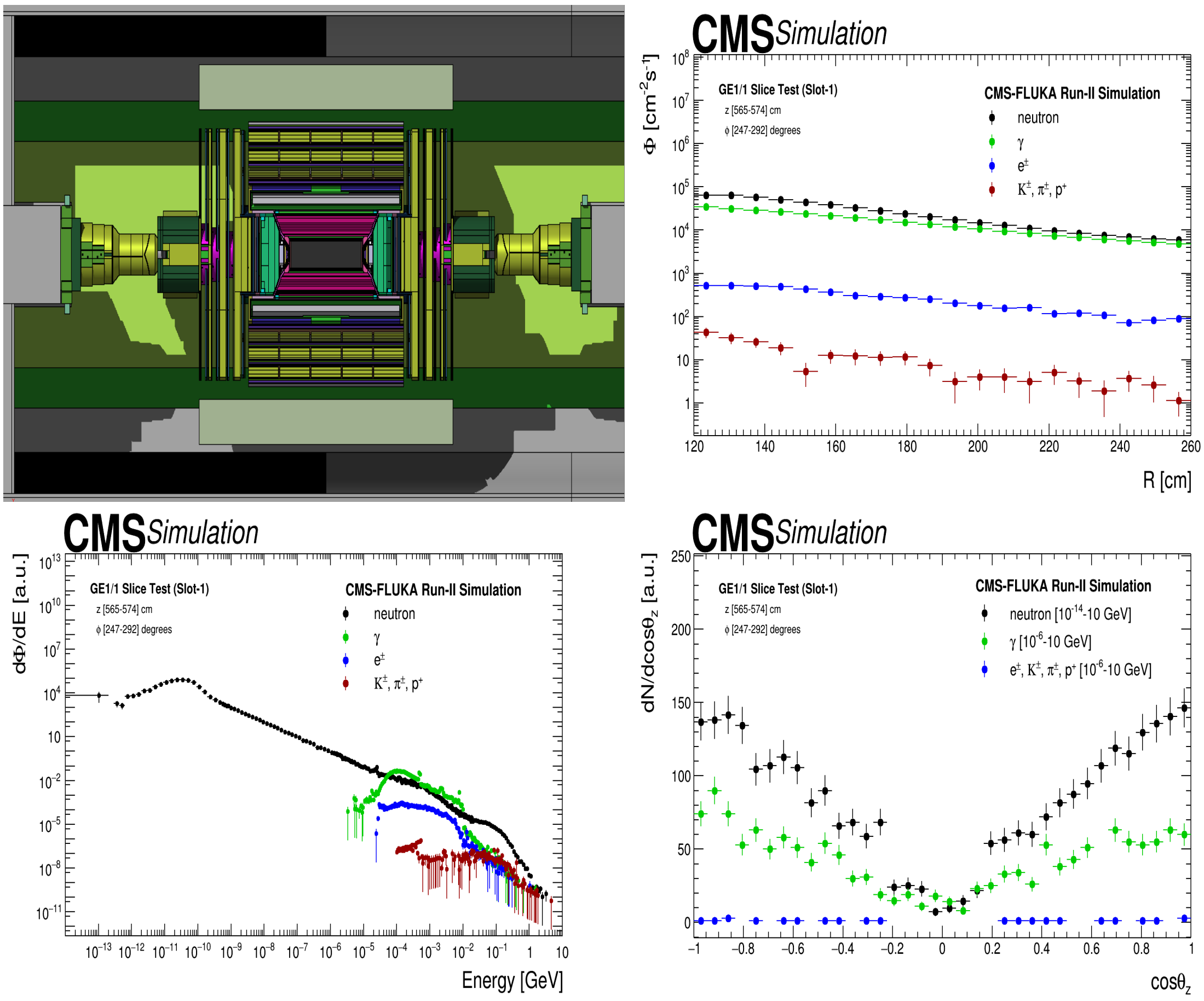

**Figure 5**. CMS geometry used in FLUKA simulation (top-left), Flux of particle arriving to GE1/1 volume, normalized to luminosity (top-right). Energy spectra of incoming particles (bottom-left), direction cosine of different particles with respect to axis perpendicular to detector surface (bottom-right)

### 3.2 Detector Response for a superchamber

The detector response to background radiation depends on the material budget and its various components. The CMS experiment has used a combination of two single trapezoidal shape triple-GEM detector combined into a so called superchamber to increase the efficiency of the hits, as two triple-GEM give hits independently. Moreover each triple-GEM detector is divided into eight $\eta$-sectors between $1.6 < |\eta| < 2.2$ [7]. The five superchambers were installed in the positions shown in figure 4. Superchambers 27, 28, 29, 30 ($\Delta\phi = 40°$) in Slot-1 are to measure muon rate, while superchamber 1 ($\Delta\phi = 10°$) in Slot-2 to test electronics and a new high voltage system [19]. These superchambers are arranged such that when all 72 superchambers are installed they will cover all $2\pi$ of each endcap wheel. The triple-GEM detector of a superchamber facing towards the interaction point of $pp$ collision is called “layer-1” while the one facing on outer side is called “layer-2”. The experimental chambers come in two flavors one “Long” and one “Short”. The dimensions are summarized in table 2 as per design requirements of the CMS experiment.

The design of triple-GEM detectors used in the CMS experiment have additional components compared to simple triple-GEM configuration described in section 2. The table 3 lists some of these important additional components and their details. The total thickness of superchamber used in the study is 73.1 mm including the space of 3.7 mm between the two chambers. Each triple GEM detector unit has GEM Electronic Board (GEB) to power the readout electronics. Every superchamber has a cooling system that consists of cooling pads and cooling pipes. The cooling pipe contains the chilled water flowing to cool the system during data taking operations. Copper is used for cooling pipe and pads due to its good thermal conductivity. The other main components of a superchamber are VFAT ASIC (Very Forward ATLAS and TOTEM ASIC) chips and Opto-Hybrid. VFATs are used for reading, digitizing and processing the signal from 384 strips from each $\eta$-sectors of the superchamber layer. The Opto-Hybrid board is plugged into GEB containing GBT (Giga-Bit Transceiver) chip sets, optical receivers and transmitters, and a FPGA (Field Programable Gate Array). Figure 6 left shows schematically various components of a single triple-GEM detector used in CMS while figure 6 right shows a Geant4 based image of a superchamber used in this study.

**Table 2**. Dimensions of "Long" and "Short" triple-GEM detector configuration in the CMS experiment.

| Configuration | Long Chamber | Short Chamber |
|---|---|---|
| y-Dimension Length | 1283.0 mm | 1135.0 mm |
| x-Dimension Shorter Length | 282.2 mm | 282.2 mm |
| x-Dimension Larger Length | 510.0 mm | 483.7 mm |

**Table 3**. Additional material, layers and their dimension used in triple-GEM detector configuration in the CMS experiment.

| Layer | z-Dimensions | Material |
|---|---|---|
| GEB | 0.1 mm/0.9 mm | Copper/FR4 |
| VFAT and Opto-Hybrid | 1.0 mm/1.6 mm | FR4/FR4 |
| Cooling Pads | 1.0 mm | Copper |
| Cooling Pipes | 8.0 mm external, 6.0 mm internal | Copper (Filled with $H_2O$) |
| Spacers | 3.0 mm/1.0 mm/2.0 mm/1.0 mm | FR4 |
| External Frame | 7.2 mm | FR4 |
| ZIG | 11.5 mm | Aluminium |
| Cover | 1.0 mm | Aluminium |

The sensitivity is calculated by simulating the detailed detector configuration as described above and is the same used in the CMS slice test [19]. Thresholds used in simulation are summarized in table 4.

**Table 4**. The energy thresholds and related parameters used in the simulation for the drift and the transfer-1 gap of triple-GEM detector configuration, assuming an effective gain of $1\times 10^4$ and a minimum charge of 3 fC in readout.

| Parameters | Drift Gap | Transfer-1 Gap |
|---|---|---|
| Minimum no. of electrons in the gap | 2 | 43 |
| Energy thresholds | 56.2 eV | 1.21 KeV |

The generation of electrons due to ionization also depends on $\langle W_i \rangle$, the effective average energy required to remove an electron, of gas mixture used in the sensitive volume of the detector. The $\langle W_i \rangle$ = 28.1 eV [20, 21] is used for threshold described in table 4 for a Ar/$CO_2$ mixture with 70/30 mixture ratio [7].

It should also be noted here that response of the single triple-GEM detector also gets affected by material of other triple-GEM detectors within the superchamber through which background particles pass and possibly interact before reaching its surface and sensitive volume.

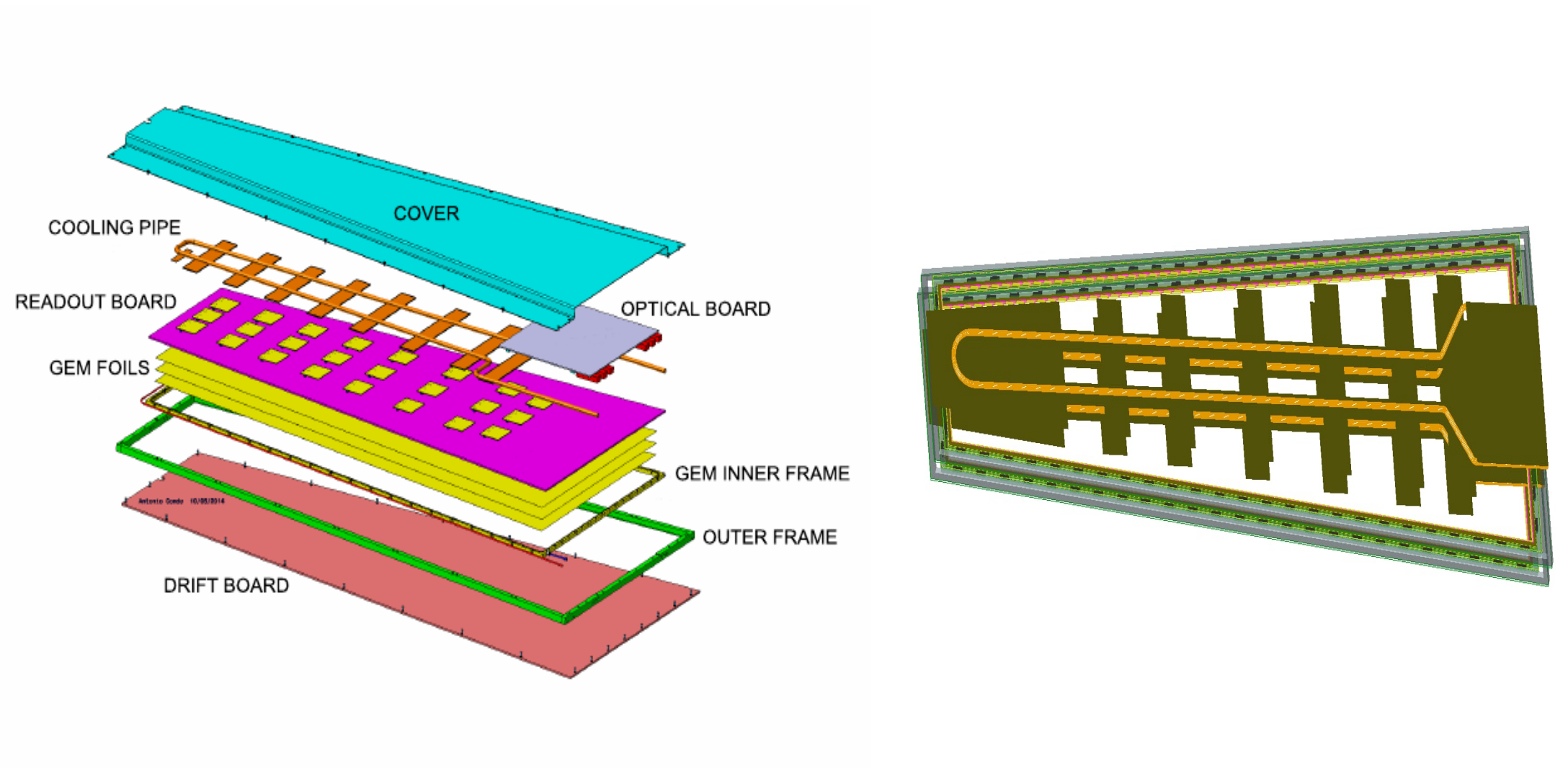

**Figure 6**. Single triple-GEM chamber with different layers (left), and Geant4 visualization of superchamber constructed by combining two single chambers (right).

## 4 Environmental background hit rates in GE1/1

The hit rate is defined as the number of detectable signals generated by particles and collected by a particular detector device per unit of time and was studied for the GE 1/1 superchambers as a function of luminosity and detector area [22]. The hit rate is one of the most important variables to measure during the calibration and monitoring. In simulation to get an estimation for hit rate the sensitivity and particle flux need to be combined as illustrated in equation 4.1.

$$\text{Hit Rate} = \sum_{type} \text{Sensitivity}(type, E, \theta) \otimes \text{Flux}(type, E, \theta, R) \tag{4.1}$$

Here *type* is the type of particle (*i.e.* neutrons, $\gamma$'s, $e^{\pm}$ and charged hadrons), $E$ is the energy of the incident particles and $\theta$ is the angle with respect to the axis perpendicular to the detector surface. The perpendicular distance from the beamline is denoted by $R$. The particle flux is obtained from FLUKA estimates presented in figure 5 top-right and multiplied by an average sensitivity. The average sensitivity is obtained by convolution of the sensitivity at a given energy and incident angle with abundance of particles at that energy and incident angle. The integration of this convolution over a given energy range gives the average sensitivity. The flux take care of all possible particles arriving from different directions at a given position at the detector while average sensitivity scales that flux into total hits at that position based on detectors response.

The simulation technique used here for sensitivity estimation is similar as described in section 2. The only main difference is that now we have a superchamber and additional material due to presence of “layer-1” in front of “layer-2” which affect layer-2’s sensitivity as its measured with respect to incident particle arriving at the superchamber surface from different directions. The sensitivity values for “layer-2” of superchamber as a function of energy of primary incident particles, arriving at superchamber surface from both sides, convoluted over all possible incident angles is shown in figure 7 for neutron, photon, electrons/positrons and charged hadrons.

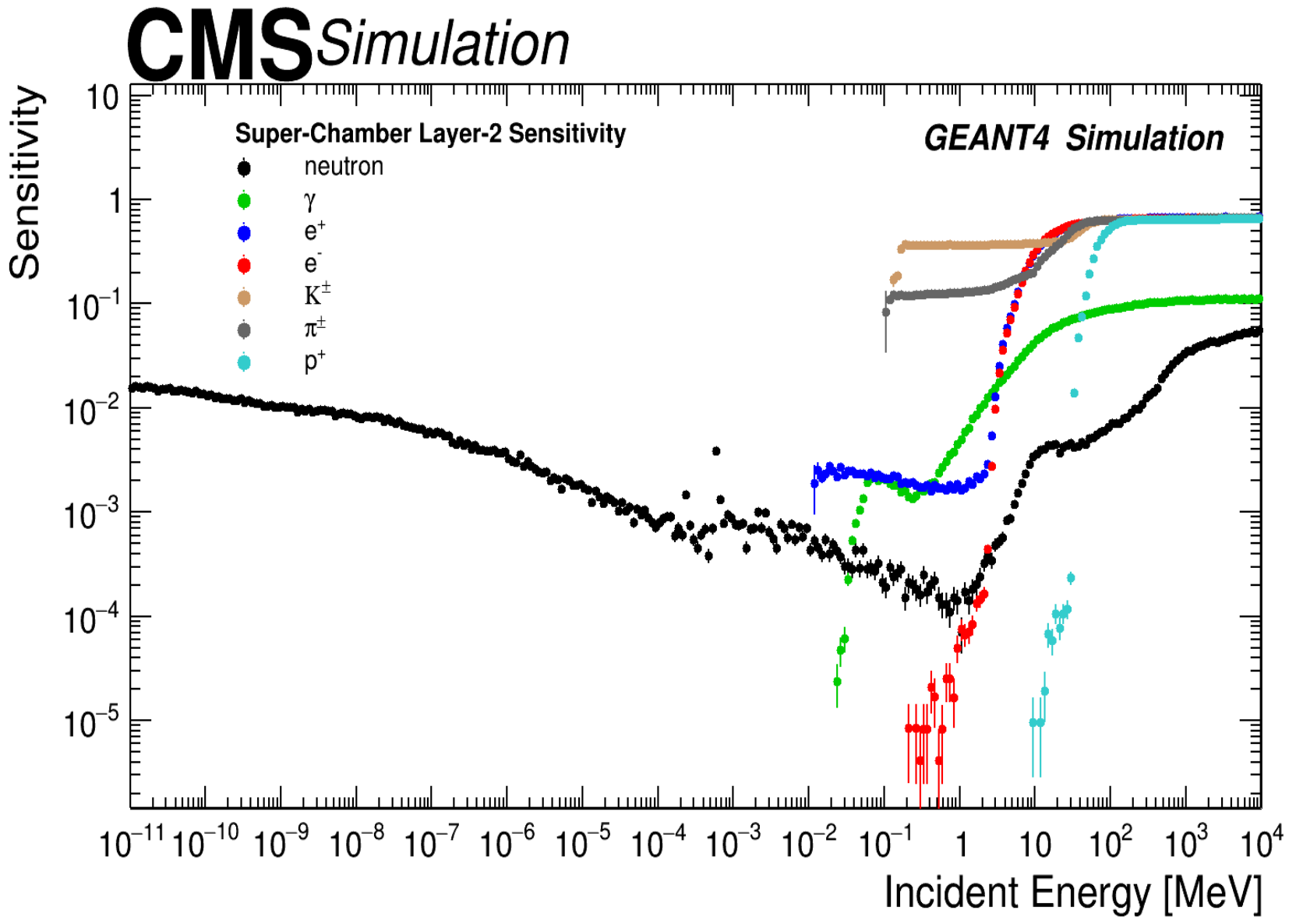

**Figure 7**. Sensitivity behavior of “layer-2” as a function of incident energy for different particles. Here the sensitivity is convoluted over all possible incident angles.

The values for the average sensitivity estimated using energy spectrum of the incident background particles and sensitivity from different energies are presented in table 5 both for "layer-1" and "layer-2". The approach of using average sensitivity is preferred rather than obtaining sensitivity for different $\eta$-sectors because these detectors have uniform response for each $\eta$-sector, as verified with X-rays during quality control tests [23].

**Table 5**. Average Sensitivity for each type of particle for the layer-1 and layer-2 of the superchamber configuration used in CMS data taking in 2018.

| Particle | Average Sensitivity of Layer-1 (%) | Average Sensitivity of Layer-2 (%) |
|---|---|---|
| Neutron | 0.64 ± 0.01 (stat.) | 0.76± 0.01 (stat.) |
| $\gamma$ | 0.28 ± 0.01 (stat.) | 0.22± 0.01 (stat.) |
| $e^{\pm}$ | 1.24 ± 0.04 (stat.) | 0.31± 0.01 (stat.) |
| Charged Hadrons ($K^{\pm}$,$\pi^{\pm}$,$p^{+}$) | 26.29 ± 1.24 (stat.) | 24.29± 1.14 (stat.) |

The effect of detector material and configuration on sensitivity could be realized from a comparison of figure 3 left and figure 7 as they represent basic geometry and CMS specific geometry respectively.

## 5 Systematics

The accuracy on the sensitivity estimations depends on a correct description of the physics processes and on a realistic detector modeling. Physics processes considered in this simulation are well known and have been validated in Geant4 framework through the years by several studies and comparisons with experimental data [24], [25]. To quantify the impact of the detector modeling on sensitivity estimations the following parameters are varied with respect to the default configuration.

- Drift Gap Width (DGW): This variable considers the possible variations in the thickness of the gas gap and GEM foils due to mechanical deformations during detector assembly. Such realistic variations [7] of ±10 % are used for Drift Gap in the simulation for both layers of the superchamber simultaneously and the impact on average sensitivity of Layer-2 is shown in table 6.

- Gas Mixture Proportion (GMP): This variable considers the possible variations in the proportions of Ar and $CO_2$ mixture during operation. Such variation were monitored during the detector operation and quality control test and found to be negligible, however a conservative variations are used in the simulation. The variation of Ar/$CO_2$ mixture considered here are (60/40) and (80/20). The variations are used in the simulation and the impact on the estimated average sensitivity of Layer-2 is shown in table 6.

The magnitude of these variations depends on the type of incident particle, kinetic energy and incident angle, and makes the detector less sensitive as compared to default configuration. The impact of these variations on sensitivity for different energy ranges are also estimated. This estimation is performed in three energy ranges for neutrons: low energy (LE) from $10^{-11}$ to $10^{-2}$ MeV, intermediate energy (IE) from $10^{-2}$ to 1 MeV, and high energy (HE) from 1 to $10^4$ MeV. Two energy ranges for $\gamma$'s and $e^{\pm}$: LE from $10^{-2}$ to 1 MeV and HE from 1 to $10^4$ MeV. The maximum variations due to GMP for neutron in LE (IE/HE) range is estimated to be 1.0 (5.6/4.6) %. For photon, $e^{\pm}$, and charged hardons these variations in LE (HE) range is estimated 0.8 (0.5) %, 2.1 (0.3) %, and 1.2 (0.1) % respectively. For DGW uncertainties similar order of variations estimated except for neutron case in IE region where low statistics dominate these variations.

The systematics arising from the method to generate a different primary source near the surface of the detector is also evaluated. The distance of source surface is varied by 2 mm on either side of the nominal value of 3 mm and average sensitivity is re-calculated to measure the impact on hit rate. A total variation of 0.5-0.8 %, 0.0-1.0 %, 0.2-0.4 %, and 3.9-8.8 % on average sensitivity is estimated for neutron, $\gamma$, $e^{\pm}$, and charged hadrons respectively. These uncertainties are also summarized in table 6.

The other systematics from simulation setup comes from the variation in the $x$-$y$ dimensions of the plane used as source of incident particles. The source plane's size, both at front and back of superchamber at a distance of 3 $mm$, are change by ±10 % and variation in average sensitivity is measured. The maximum variation for a 10 % larger size of source plane is found to be about 1.3 % or less for any given type of incident particle. For a 10 % smaller size of source surface give a maximum variation of 6.7 % or less.

**Table 6**. Variations on simulation parameters and their impact on estimated average sensitivity, the numbers were obtained comparing with average sensitivity of the layer-2 as shown in table 5.

| Parameters | Values for Variations | Impact on average sensitivity of Layer-2 (in %) | | | |
|---|---|---|---|---|---|
| | | Neutron | $\gamma$ | $e^{\pm}$ | Charged Hadrons |
| DGW | Drift Gap: 2.7 mm | 0.6 | 1.4 | 1.0 | 0.5 |
| | Drift Gap: 3.3 mm | 0.3 | 0.4 | 0.0 | 0.6 |
| GMP | $Ar/CO_2$ (60/40) | 1.0 | 0.4 | 0.3 | 0.5 |
| | $Ar/CO_2$ (80/20) | 0.1 | 0.9 | 0.9 | 0.2 |
| source $z$-position | 1 mm | 0.6 | 0.4 | 0.9 | 4.0 |
| from superchamber | 5 mm | 1.0 | 1.4 | 0.9 | 9.9 |

The uncertainties described in table 6 for different types of incident particles directly translate to uncertainty on the hit rate as per their relative contribution to total hit rate. The final uncertainties on hit rate is dominated by neutron and photon contributions.

The systematic uncertainty associated with the particle flux is estimated comparing two CMS-FLUKA geometry scenarios both of them consistent with the Run-2 data taking scenario, the uncertainty was evaluated comparing the particle flux between the two scenarios and the result is a variation of 10 to 20 % as a function of R with a mean value of 15 %, the latest is considered as the associated uncertainty.

The final combined uncertainty from FLUKA simulation and Geant4 detector simulation, including various parameters, is estimated to be about ~14.5 % on hit rate, after adding them in quadrature for the whole range of R considered in this study.

## 6 Comparison of background modeling and Experimental data

A comparison between simulation and experimental measurements can be used to validate the model presented in this study. Figure 8 compares the measured hit rate from data of layer-2 of superchamber 28 from CMS experiment with the prediction of simulation. The simulation results are obtained by taking the convolution of average sensitivity shown in table 5 and values shown in figure 5 top-right. The different data points correspond to different radial distance from the beamline i.e, $z$-axis, to the centre of the different $\eta$-sectors. As expected the hit rate tends to be higher in lower the $R$ (higher $\eta$) region as flux from collision is primarily directed towards this region. The bottom plot of figure 8 shows the ratio of data and simulation along with the estimated systematic uncertainty. The contributions from different types of particle to total hit rate is shown in figure 9 for layer-2. The largest contribution comes from neutrons while photons contributes about ~ 15 %. Charged hadrons and $e^{\pm}$ contributes about 1% only.

The simulation model from FLUKA+Geant4 discussed in this paper describes the data hit rate well within the total uncertainties, shown as orange and blue band around simulation prediction. The total uncertainty includes both statistical uncertainty and all the systematics described in previous section added in quadrature. The uncertainty on data for hit rate includes the statistical uncertainty and uncertainties on the luminosity measurement.

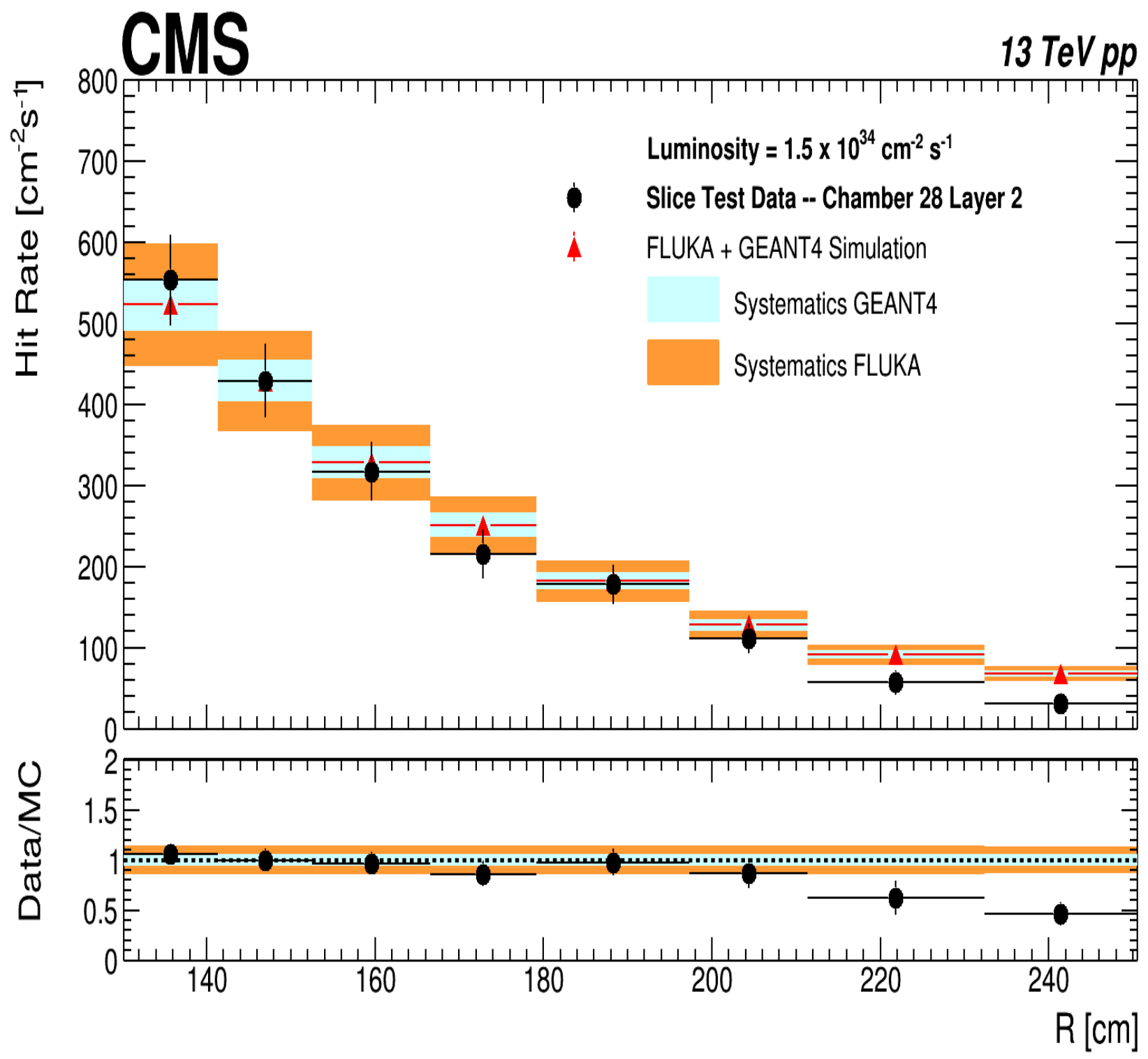

**Figure 8**. Comparison of slice test data with FLUKA+Geant4 Simulation for Layer-2 of superchamber No. 28. The systematic uncertainties from Geant4 simulation method and detector parameters are shown as shaded blue band around the simulation prediction. The uncertainty from FLUKA simulation are shown as orange band after adding them in quadrature with systematics shown in blue band. The bottom panel shows the ratio of data and simulation prediction.

The simulation prediction for first two $\eta$ sectors show a larger discrepancy with respect to data most likely due to a higher threshold set during data taking operations, such effects are difficult to describe with current simulation. In future study the framework can be improved by adding a detailed description of the electron multiplication (avalanche) using GARFIELD software and a simulation of the electronics.

## 7 Summary

Environmental background hit rates on the CMS triple-GEM chamber in pp collisions at the LHC are evaluated by modeling radiation environment and detector response using a framework of FLUKA and Geant4 simulation packages. FLUKA simulation provides incident particle's kinetic energy and angle in radiation environment. Geant4 simulation models particle interactions based on accurate material description of GEM chambers with an operation condition. The hit rates are obtained by combining sensitivity and particle flux, and compared with measurements at a luminosity of $1.5 \times 10^{34}$ cm$^{-2}$ s$^{-1}$ at 13 TeV. The predicted hit rates and experimental data agree within its uncertainties. This modeling is generic, so it can be used for evaluation of hit rates on other detectors at higher luminosities, providing relevant information for detector design and operation at High-Luminosity LHC.

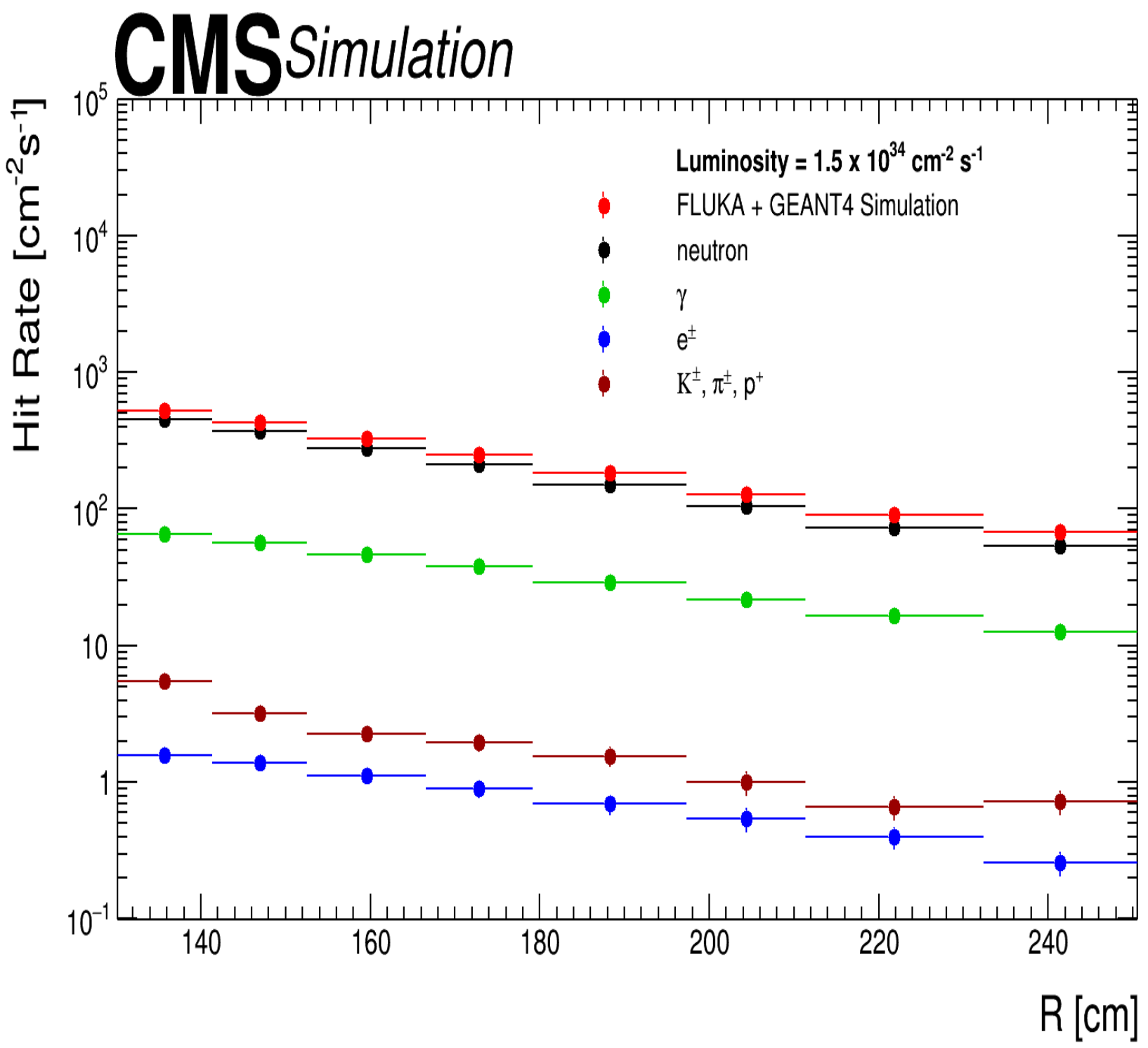

**Figure 9**. Prediction from simulation for hit rate contribution from different background particles at layer-2 of superchamber.

## Acknowledgments

We gratefully acknowledge support from FRS-FNRS (Belgium), FWO-Flanders (Belgium), BSFMES (Bulgaria), MOST and NSFC (China), BMBF (Germany), CSIR (India), DAE (India), DST (India), UGC (India), INFN (Italy), NRF (Korea), CONACYT (Mexico), MoSTR (Sri Lanka), DOE (U.S.A.), and NSF (U.S.A.).